\documentclass[prb,twocolumn,showpacs,floatfix,amsmath,amssymb,superscriptaddress]{revtex4}
\usepackage[dvips]{graphicx}
\usepackage{latexsym}
\usepackage{graphicx}
\usepackage{times}
\usepackage{amsmath}
\usepackage{dcolumn}
\usepackage{latexsym,amsmath,amssymb,bm,euscript}
\begin{document}

\title{Interplay between lattice-scale physics and the quantum Hall
  effect in graphene}

\author{Jason Alicea}
\affiliation{Physics Department, University of California, 
Santa Barbara, CA 93106}
\author{Matthew P. A. Fisher}
\affiliation{Microsoft Research, Station Q, University of California, Santa Barbara, CA 93106}

\date{\today}

\begin{abstract}

Graphene's honeycomb lattice structure underlies much of the
remarkable physics inherent in this material, most strikingly through
the formation of two ``flavors'' of Dirac cones for each spin.  In the
quantum Hall regime, the resulting flavor degree of freedom leads to
an interesting problem when a Landau level is partially
occupied.  Namely, while Zeeman splitting clearly favors
polarizing spins along the field, precisely how the states for each
flavor are occupied can become quite delicate.  Here we focus on clean
graphene sheets in the regime of quantum Hall ferromagnetism, and discuss how 
subtler lattice-scale
physics, arising either from interactions or disorder, 
resolves this ambiguity to measurable consequence.  Interestingly, 
such lattice-scale physics 
favors microscopic symmetry-breaking order coexisting with the usual 
liquid-like quantum Hall physics emerging on long length scales.  The
current experimental situation is briefly reviewed in light of
our discussion.

\end{abstract}

\maketitle


\section{Introduction}

Whereas electrons confined to 2D in 
conventional GaAs heterostructures are well-represented as residing in
a continuous space, the behavior of electrons in graphene is
inextricably tied to 
their underlying honeycomb lattice structure.  This simple fact is
central to much of the exceptional physics in the graphene 
quantum Hall effect.  Most
dramatically, in zero field the honeycomb band structure supports 
two Dirac cones centered
at momenta $\pm {\bf Q} = \pm(4\pi/3,0)$, so that graphene 
realizes a ``relativistic'' dispersion (over a 
significant energy range).
In the presence of a perpendicular magnetic field, the Dirac spectrum
develops into Landau levels that are approximately four-fold
degenerate, one factor of two coming from spin and the other from the 
two ``flavors'' of Dirac cones.
Taking into account
this degeneracy as well as a Berry phase acquired from the cyclotron
motion leads to the prediction of quantized Hall plateaus\cite{Zheng, 
Gusynin, Peres1,Peres2} with
$\sigma_{xy} = \nu e^2/h = 4(j+1/2)e^2/h$, whose
observation was one of the early achievements in the study of
graphene \cite{IQHE1,IQHE2}.  

The Landau level structure as well as the formation of these
plateaus can be readily captured by a continuum Dirac equation with SU(4)
symmetry (reflecting invariance under spin and flavor rotations), 
with no further considerations
of graphene's lattice structure.  In a sense, the problem then becomes
rather similar to GaAs 2D electron systems, with a modified spectrum.
But is this the full story?  Simple symmetry considerations suggest that
it is not.  The approximate four-fold Landau level degeneracy noted
above is clearly broken down by Zeeman coupling, but this still leaves
an SU(2) flavor symmetry in the continuum theory.  This residual
symmetry can certainly be \emph{spontaneously} broken, as occurs due
to Coulomb exchange in quantum Hall ferromagnetism wherein plateaus
occur at \emph{all} integer filling factors (see the 
reviews of Refs.\ \onlinecite{GirvinReview, KunYangReview}).  
However, it is important to observe
that there is no microscopic reason for its existence in the first
place.  This suggests that \emph{something} can, and will, 
\emph{explicitly} lift this degeneracy further.
Such effects were absent in initial experiments, due presumably to
their being overshadowed by disorder.  
Given the observation of additional $\nu = 0, \pm 1$, and $\pm 4$ 
plateaus at higher magnetic fields \cite{HighFieldExpt,HighFieldExpt2} 
as well as
the inevitable future enhancement of sample quality, the question of
precisely how the flavor degeneracy is lifted in graphene quantum 
Hall states is not only interesting, but also relevant.  Such issues
are important to resolve for a number of reasons.  For instance, what 
is the precise nature of these new integer quantum Hall states?
Looking forward, what type of states should be anticipated in the fractional
regime?  How do the low-lying excitations behave?  What, if any,
symmetries are broken in such states?  And how can such properties be
detected?  

The central goal of this short review is to demonstrate that subtler
lattice-scale physics provides such a degeneracy-lifting mechanism.
There are a number of possible sources for this physics, which have
been briefly summarized in Ref.\ \onlinecite{KunYangReview}.  Here we will
expand on some of these ideas, concentrating on clean graphene systems
in the regime of quantum Hall ferromagnetism, which we believe to be
of current experimental relevance.   
For the most part, we will examine a continuum Dirac theory 
derived from a simple lattice Hamiltonian that incorporates the 
electron kinetic energy and Coulomb repulsion, and
demonstrate that such a minimal model indeed propagates
important flavor-symmetry-breaking interactions into the continuum.  
Interestingly, this model predicts that quantum Hall states at filling
factors $\nu = \pm
1$ are not liquid-like at all on the lattice scale, but rather exhibit
microscopic charge density wave order.  
Similarly, a broken-symmetry pattern with a tripled unit cell is
favored at other odd-integer $\nu$.  (See Fig.\ \ref{Orders}.)
We then briefly summarize a recent numerical study of such an
interacting lattice model \cite{ShengNumerics}, 
which confirms these conclusions in
the clean limit and extends the analysis by incorporating disorder.  
An alternate
degeneracy-lifting mechanisms originating from disorder-driven
variations in the electron hopping strengths as proposed by Abanin
\emph{et al}.\ \cite{AbaninDisorder} is also reviewed.  
Finally, we conclude with a
brief assessment of the current experimental situation regarding these
subtle yet interesting lattice-scale effects.

\section{Continuum Theory from an Interacting Lattice Model}

A minimal lattice model for graphene that can be expected to realize
integer and fractional quantum Hall states when a strong magnetic
field is turned on should consist of two parts: electron kinetic 
energy plus Coulomb repulsion.  It is reasonable, then, to start by
understanding the physics contained in such a Hamiltonian, which we
write using second quantization as 
\begin{eqnarray}
  H &=& H_{\rm KE} + H_{\rm Coul},
\label{latticeH}
\\
  H_{\rm KE} &=& -t\sum_{\langle {\bf x x'}\rangle}\sum_{\alpha =
  \uparrow,\downarrow} [c^\dagger_{\alpha {\bf x}}
  c_{\alpha {\bf x'}} + {\rm H.c.}],
\\
  H_{\rm Coul} &=& U \sum_{\bf x}\bigg{[}\frac{1}{4}(n_{\bf x})^2
  -\frac{1}{3}{\bf S}({\bf x})^2\bigg{]}
\nonumber \\
  &+& \frac{1}{2}\sum_{\bf x \neq x'} V({\bf x -
  x'})n_{\bf x}n_{\bf x'}.
  \label{Hcoul}
\end{eqnarray}
In these expressions, $c^\dagger_{\alpha{\bf x}}$ is the electron
creation operator for spin $\alpha$, $n_{\alpha{\bf x}} =
c^\dagger_{\alpha{\bf x}}c_{\alpha{\bf x}}$ is the
corresponding electron number operator, 
$n_{\bf x} = n_{\uparrow {\bf x}} + n_{\downarrow {\bf x}}$, 
and ${\bf S}({\bf x}) = \frac{1}{2}
c^\dagger_{\alpha {\bf x}}{\bm \sigma}_{\alpha\beta} c_{\beta {\bf
x}}$ is the usual spin operator with ${\bm \sigma}$ a vector of Pauli
matrices.  The first term in $H_{\rm Coul}$ contains the on-site
repulsion energy, written in a manifestly SU(2)-invariant form, while
the second represents the long-range part of the Coulomb repulsion,
with $V({\bf x})$ the Coulomb potential.  
Normal ordering of the operators in Eq.\ (\ref{Hcoul}) is understood.

The kinetic energy term in the Hamiltonian can be readily diagonalized
to reveal two Dirac points at momenta $\pm {\bf Q} = \pm(4\pi/3,0)$,
which lie at the Fermi level when there is one electron per site.  
To describe excitations near the two nodes, it is highly convenient to
pass to a continuum formulation by expanding the momentum-space fermion 
operators in terms of two flavors
of continuum Dirac fermion fields (denoted $R$ and $L$) as follows,
\begin{eqnarray}
  c_{\alpha {\bf q+Q} a} &\sim& \gamma \psi_{\alpha R a}({\bf q})
  \label{DiracFields1}
  \\
  c_{\alpha {\bf q-Q} a} &\sim&  \gamma i \eta^y_{ab} \psi_{\alpha L b}({\bf
  q}),
  \label{DiracFields2}
\end{eqnarray}
where $\gamma = \sqrt{2}/(3^{1/4}l)$ ($l$ is the linear system size).  
Here and below we reserve indices $\alpha,\beta$ for spin, $A,
B$ for flavor, and $a,b$ for the honeycomb sublattice.  
It is also useful to employ Pauli matrices
$\sigma^j_{\alpha \beta}$, $\tau^j_{AB}$, and $\eta^j_{ab}$ 
that contract with the spin, flavor, and sublattice indices,
respectively.  We will use the convention that suppressed 
indices on the fields are implicitly summed
(\emph{i.e.}, $\psi^\dagger \psi \equiv \sum_{\alpha A
  a}\psi^\dagger_{\alpha A a } \psi_{\alpha A a}$).

With these identifications, it is straightforward to
recast the full interacting Hamiltonian into a continuum Dirac theory.
Upon turning on the external magnetic field, one can essentially guess
the leading kinetic energy and interaction terms, namely
\begin{eqnarray}
  {\mathcal H}_0 &=& -i \hbar v \int d^2{\bf x}
  \psi^\dagger [\eta^x D_x+\eta^y D_y]\psi 
  \nonumber \\
  &+& \frac{1}{2} \int d^2{\bf x}d^2{\bf x'} 
  \rho_{\rm tot}({\bf x}) V({\bf x-x'}) \rho_{\rm tot}({\bf x'}),
  \label{H0}
\end{eqnarray}
with $v \approx 10^6$ m/s the Fermi velocity, 
$D_j = \partial_j - i (e/\hbar) A_j$, and $\rho_{\rm tot} =
\psi^\dagger \psi$ the total density.  For concreteness we take the
magnetic field ${\bf B} = \nabla \times {\bf A}$ 
in the $+{\hat {\bf z}}$ direction, normal to the
graphene sheet.  These contributions clearly
exhibit an SU(4) symmetry, being invariant under arbitrary spin/flavor
rotations.  A more careful derivation, however, reveals a number of
anisotropy terms, which break this SU(4) symmetry down to 
${\rm U}(1)_{\rm spin} \times[{\rm U}(1) \times {\rm Z}_2]_{\rm
  flavor}$.  Writing the full Hamiltonian as ${\mathcal H} = {\mathcal
  H}_0 + {\mathcal H}_1$, these additional interactions can be written
as follows,
\begin{eqnarray}
  {\mathcal H}_{1} &=& \int d^2 {\bf x} \big{\{}- g \mu_B {\bf
  B}\cdot {\bf S}_{\rm tot}
  + \frac{1}{4}u_0 [\rho_{\rm tot}^2 
  +\rho_{\rm stag}^2
  \nonumber \\
  &-& \frac{8}{3}({\bf S}_{R1}^2+{\bf S}_{L2}^2 + 6 {\bf
  S}_{R1}\cdot {\bf S}_{L2} + (1\leftrightarrow 2)) ]
  \nonumber \\
  &-& \sum_{\bf r} v_1({\bf r}) 
  \rho_{\rm stag}({\bf x+r})\rho_{\rm stag}({\bf x}) 
  \nonumber \\
  &-& u_2 [J_+^\dagger J_+ + J_-^\dagger J_-]\big{\}},
  \label{H1}
\end{eqnarray}
where ${\bf S}_{\rm tot} =
\frac{1}{2} \psi^\dagger {\bm \sigma}\psi$ is the uniform spin 
density,
$\rho_{\rm stag} = \psi^\dagger \tau^z \eta^z \psi$ represents the
staggered electron density between sublattices 1 and 2 of the
honeycomb (see Fig.\ \ref{Orders}), 
${\bf S}_{Aa} = \frac{1}{2} \psi^\dagger_{Aa} {\bm \sigma} \psi_{Aa}$,
and $J_+ = \psi^\dagger_{R1} \psi_{L2}$ and $J_- =
\psi^\dagger_{R2}\psi_{L1}$ represent components of the density
oscillating at wave vectors $\pm 2{\bf Q}$.  
The sum in Eq.\ (\ref{H1}) is over Bravais lattice vectors.  Once
again, normal ordering is understood in the interactions above.

While the presence of these short-range anisotropy terms is
not \emph{a priori} obvious (apart from the Zeeman coupling), each
has a simple physical interpretation.  
The second term in Eq.\ (\ref{H1}) merely encodes the on-site $U$
repulsion; displaying the lattice constant $a_0$ explicitly, we have
\begin{equation}
  u_0 = \sqrt{3} a_0^2 U/4.  
  \label{u0}
\end{equation}
The third, whose coupling is given by
\begin{equation}
  v_{1}({\bf r}) = \frac{\sqrt{3}a_0^2}{8} [ V({\bf r}
  + a_0/\sqrt{3}{\bf \hat y}) - (1-\delta_{{\bf r},{\bf 0}})V({\bf r})],
  \label{v1}
\end{equation}
represents an inter-sublattice repulsion that 
reflects the smaller Coulomb energy cost
for electrons residing on the same sublattice versus opposite 
sublattices.  While this interaction is certainly short-ranged, the 
above non-local form must be retained when dealing
with filling factors $\nu = \pm 1$ to have an effect.  At other
integer filling factors, however, a purely local form suffices, and
one can replace
\begin{eqnarray}
  &&\sum_{\bf r}v_1({\bf r}) \rho_{\rm stag}({\bf x}+{\bf r})\rho_{\rm
  stag}({\bf x}) \rightarrow u_1 \rho_{\rm stag}^2,
\\
  &&u_1 \approx 
  \frac{1}{\sqrt{3}}a_0^2\bigg{(}\frac{e^2}{4\pi\epsilon a_0}\bigg{)},
\end{eqnarray}
where $\epsilon$ is the (unscreened) dielectric constant.
The final term represents the intra-sublattice repulsion between 
density components oscillating at ${\pm 2{\bf Q}}$, with
\begin{equation}
  u_2 \approx \frac{4}{\sqrt{3}}u_1.
  \label{u2}
\end{equation}
We note that the $u_2$ interaction was
also noticed by Goerbig \emph{et al}., \cite{Goerbig} though at odd-integer
filling factors we find that the $v_1$ term provides the leading
flavor-symmetry-breaking interaction.  

Before turning to the implications of these interactions 
for quantum Hall states, it is useful to note the hierarchy of 
energy scales in the problem.  The kinetic energy term in the
SU(4)-invariant ${\mathcal H}_0$ gives rise to four-fold degenerate
Landau levels\cite{Zheng, Haldane} indexed by an integer $n$, with energies 
$E_n = {\rm sign}(n)\sqrt{2e \hbar v^2 B|n|}$.  The spacing between
the $n = 0$ and $n = \pm 1$ levels is the largest energy scale at
roughly $400 \sqrt{B[{\rm T}]}$ K, where $B[{\rm T}]$ is the magnetic
field evaluated in Teslas.  The characteristic Coulomb energy scales
with the field in the same way:
\begin{equation}
  {\mathcal E}_C \equiv \frac{e^2}{4\pi \epsilon_{\rm RPA} \ell_B} \sim 100 
  \sqrt{B[{\rm T}]}~{\rm K},
  \label{Ec}
\end{equation}
where $\ell_B$ is the magnetic length and 
$\epsilon_{\rm RPA}$ is the screened dielectric constant
computed within the random phase approximation \cite{Gonzalez}, which yields
$\epsilon_{\rm RPA} \approx 5 \epsilon_0$.  The energies
associated with the symmetry-breaking terms in ${\mathcal H}_1$ are
much smaller at laboratory fields.  The Zeeman energy for
instance is $g\mu_B B \sim B[{\rm T}]$ K.  Furthermore, as emphasized in Ref.\
\onlinecite{Goerbig} the energies for the short-range interactions 
in ${\mathcal H}_1$ are down by factors of $a_0/\ell_B$ compared 
with ${\mathcal E}_C$.  
The associated scale for these terms is 
\begin{equation}
  {\mathcal E}_{\rm short-range} \equiv \frac{e^2}{4\pi \epsilon a_0}
  \bigg{(}\frac{a_0}{\ell_B}\bigg{)}^2 \sim B[{\rm T}]~{\rm K} ,
  \label{Esr}
\end{equation}
the first factor being the
characteristic lattice-scale Coulomb energy and the second reflecting the
average number of electrons in the highest occupied Landau level 
per unit cell.  As we will discuss below, an 
exception occurs at filling factors $\nu = \pm 1$, where the scale for
flavor symmetry breaking is down by an additional factor of
$a_0/\ell_B$.  Given the clear separation in energy scales above, it
is reasonable to first consider the physics contained in
${\mathcal H}_0$, and then take into account the additional
symmetry-breaking terms in ${\mathcal H}_1$.  This is the strategy we
will follow below.

\section{Implications for quantum Hall states}

Let us now very briefly review some aspects of integer quantum Hall physics 
expected from the SU(4)
invariant Hamiltonian ${\mathcal H}_0$.  
Apart from the integer quantum Hall states occurring at filling
factors $\nu = 4(j+1/2)$, where the highest occupied Landau level is
completely full, it is well-established that 
quantum Hall ferromagnetism emerges at other integer filling
factors due to Coulomb exchange
\cite{GirvinReview,GrapheneMacDonald,KunYangReview}.  Here, in the
absence of anisotropy terms the SU(4) symmetry enjoyed by ${\mathcal
  H}_0$ is broken spontaneously, giving way to gapless Goldstone modes
analogous to spin waves\cite{KunYangSkyrmions} (the number depends on
$\nu$) and SU(4) 
skyrmions\cite{Skyrmions, Arovas} that provide the
lowest-energy charge excitations in the $n = 0,\ldots,\pm 3$ levels
\cite{KunYangSkyrmions}.  

The physics of quantum Hall ferromagnetism in graphene is substantially 
modified by the anisotropy terms in
${\mathcal H}_1$.  Only a subset of the SU(4) symmetry will be spontaneously
broken, the Goldstone modes will generally
become gapped, and the skyrmion character will change as well.  Here we will
focus for the most part on elucidating the qualitative features of
these effects, obtained within a Hartree-Fock framework; 
quantitative aspects have been discussed elsewhere
\cite{AliceaGraphene, KunYangSkyrmions, GrapheneBosonization}.  
We will specialize to the $n = 0$ and $n =
1$ Landau levels ($n = -1$ is related by particle-hole symmetry).  
To this end, we first note a simple yet crucial 
feature of the Landau level wave functions that will prove illuminating.  
Namely, the $n = 0$ 
single-particle wave functions for flavor $L$ reside 
entirely on honeycomb sublattice 1, while flavor $R$ states
reside entirely on sublattice 2.  In contrast, 
the probability weight for $n \neq 0$ wave functions is evenly
distributed between both sublattices.  The effect of ${\mathcal H}_1$
in these two cases differs completely as a result of this
distinction.  

\begin{figure} 
  \begin{center} 
    {\resizebox{6cm}{!}{\includegraphics{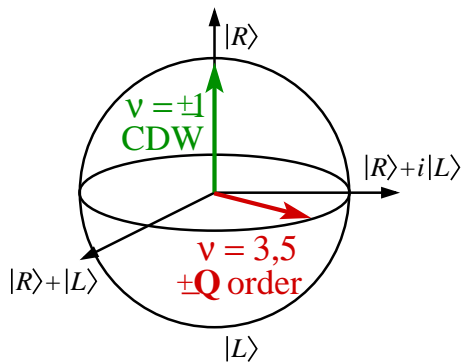}}} 
  \end{center} 
  \caption{Order-parameter space for flavor-polarized 
  states.  Interactions favor ``easy-axis'' polarization at 
  $\nu = \pm 1$, but ``easy-plane'' polarization at $\nu = 3,5$.  The
  corresponding lattice-scale order is shown schematically in Fig.\
  \ref{Orders}.  Disorder of the type discussed in Sec.\ V, however,
  favors easy-plane polarization at $\nu = \pm 1$.}
  \label{FlavorSphere} 
\end{figure}

\begin{figure} 
  \begin{center} 
    {\resizebox{8cm}{!}{\includegraphics{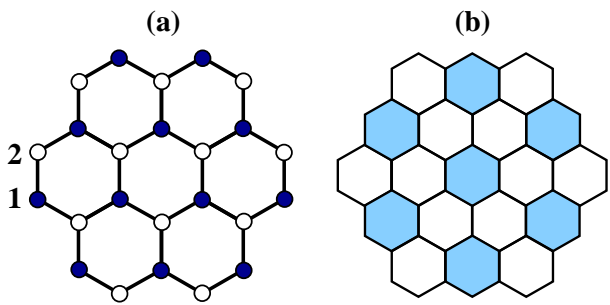}}} 
  \end{center} 
  \caption{Lattice-scale symmetry-breaking patterns favored by interactions 
  at (a) $\nu = \pm 1$ and (b) $\nu = 3,5$.  On the left, the electron
  density is higher on one of the two sublattices.  On the right
  translation symmetry is broken, along with, in general, rotation and
  reflection symmetries due to the formation of lattice-scale currents
  circulating around hexagons.  For example, one such ordering
  involves currents circulating uniformly around the shaded hexagons.} 
  \label{Orders} 
\end{figure}

Consider the $n = 0$ Landau level first.  Employing the standard
Landau level projection, the interactions simplify
greatly here due to the character of the
wave functions noted above, with the $u_2$ term dropping out entirely.  
At filling factor $\nu = -1$, corresponding to a quarter-filled level, 
Zeeman coupling clearly favors a spin-polarized state;
the on-site $U$ interaction can then play no role due to Pauli
exclusion.  The fate of the flavor degree of freedom is thus set by
the sublattice repulsion in ${\mathcal H}_1$.  It is useful
to note that a general flavor-polarized state can be expressed as
\begin{equation}
  |\theta,\phi \rangle = \prod_{m =
   0}^\infty\bigg{[}\cos\frac{\theta}{2}
   c^\dagger_{\uparrow R,m} + \sin\frac{\theta}{2}
   e^{i\phi}c^\dagger_{\uparrow L,m}\bigg{]}|{\rm vac}\rangle,
  \label{FPstate}
\end{equation}
where $c^\dagger_{\uparrow A,m}$ adds a spin-up, flavor-$A$ particle with 
angular momentum $m$ into the $n = 0$ Landau level and $|{\rm
vac}\rangle$ corresponds to the empty $n = 0$ level.  The angles
$\theta,\phi$ specify the polarization direction in ``flavor space'',
shown schematically in Fig.\ \ref{FlavorSphere}.
In particular, ``easy-axis'' flavor polarization is favored here,
with $\theta = 0$ or $\pi$, resulting in a state with all the
electrons occupying the $n = 0$ level spontaneously choosing to reside
on one honeycomb lattice as shown in Fig.\ \ref{Orders}(a).  
The physics behind this microscopic
charge density wave (CDW) is simple: the electrons can stay farther apart
from one another by remaining on one sublattice.  

It is important to note that this effect is rather weak.  First, the 
gap $\Delta_f$ for exciting ``flavor
waves'' (which weakly modulate the system away from the perfect CDW, 
much like a spin wave) is quite small: 
\begin{equation}
  \Delta_{f} \approx 4 \times 10^{-3} (B[{\rm T}])^{3/2} {\rm K}, 
\end{equation} 
which is down by an extra small factor
of $a_0/\ell_B$ compared to ${\mathcal E}_{\rm short-range}$ in Eq.\
(\ref{Esr}).  Despite appearances, a local
$\rho_{\rm stag}^2$ interaction is insensitive to
$\theta$ and $\phi$ above, for the same reason that $(S^z)^2$ is
trivial for a spin-1/2 moment.  Rather, it is the non-local part of
the sublattice repulsion in ${\mathcal H}_1$ that provides the
degeneracy lifting, resulting in the additional small factor noted above.  
The spin-wave gap, by contrast, is much larger at
$\Delta_s = g\mu_B B \approx B[{\rm T}]$ K; hybrid ``spin-flavor
waves'' cost a similar energy.
The smallness of the flavor anisotropy implies that ``flavor
skyrmions'' will provide the lowest-energy charge excitations, but 
also suggests that other 
mechanisms not included here may influence the flavor degree of
freedom, as we will discuss below.  Second, only the relatively small
number of electrons in the $n = 0$ level participate in the CDW
(around $1.4 \times 10^{-5}B[{\rm T}]$ electrons per hexagon),
though in principle its presence may be detected with STM or NMR
measurements \cite{AliceaGraphene}.  
Analogous results hold for the particle-hole
related $\nu = 1$ filling factor.  

The ground state at $\nu = 0$ depends on the competition between
Zeeman coupling and the on-site $U$ versus the sublattice repulsion,
which is much more effective here since even a local $\rho_{\rm
  stag}^2$ interaction behaves nontrivially on polarized states.
If the former dominate, then the ground state will clearly be 
spin-polarized, with electrons equally occupying both sublattices.  
This transpires provided $g\mu_B B + 2u_0\rho_0 > 4u_1 \rho_0$, where
$\rho_0 = 1/(2\pi\ell_B^2)$ is the density of a quarter-filled Landau
level.  For clean graphene sheets on a SiO$_2$ substrate, 
we estimate that this
inequality is satisfied if $U \gtrsim 4$ eV.  Otherwise the sublattice
repulsion will dominate, and a spin-singlet CDW emerges.  We emphasize
that here the CDW order can be much more robust than at $\nu = \pm 1$,
the flavor-wave gap now being set by $4u_1 \rho_0-(g\mu_B B +
2u_0\rho_0)$ which should generically be of order ${\mathcal E}_{\rm
  short-range}$.
Which scenario prevails is unclear 
because of uncertainties
in the magnitude of $U$, \cite{Uestimate} though a spin-polarized state can
always be achieved by tilting the field to enhance the effective 
$g$-factor.  In either case, the charge gap is expected to be set by
mixed spin/flavor-textured skyrmions.  For instance, in the
spin-polarized state skyrmions will have lower energy if they degrade
the spin polarization while simultaneously restoring CDW order.  
The dependence of the activation energy on
an in-plane field can allow one to distinguish between these ground
states.   An increase is expected in the spin-polarized state since the
larger effective $g$-factor reduces the optimal 
skyrmion size and hence raises their energy, while a
decrease is expected for the spin-singlet CDW since here skyrmions
induce a revival of spin polarization.  

Next we discuss the $n = 1$ Landau level, beginning with filling
factor $\nu = 3$.  The physics is similar for $\nu = 5$ and will not
be discussed separately.  Since the wave functions now live equally 
on both sublattices, one must wrestle with all terms in 
${\mathcal H}_1$.  Satisfying the Zeeman coupling requires full
spin polarization, so again the on-site $U$ repulsion drops out.  
A general flavor-polarized state can be written as in
Eq.\ (\ref{FPstate}), with the creation operators now filling $n = 1$
Landau levels.  Both the $u_1$ and $u_2$ terms lift the ordering
degeneracy here, though the effects are subtle since the form of the
wave functions dictates that $\langle \rho_{\rm stag}\rangle = \langle
J_{\pm}\rangle = 0$ for any $\theta, \phi$.  Rather, the
symmetry-breaking comes from exchange contributions.  In particular,
$u_1$ favors ``easy-plane'' polarization, \emph{i.e.}, $\theta =
\pi/2$ and arbitrary $\phi$, while $u_2$ favors easy-axis
polarization.  A quantitative estimate suggests that the former
easy-plane state, which is characterized by an order parameter 
$\langle \psi^\dagger (\tau^x+i \tau^y)\psi\rangle\neq 0$, has lower
energy.  This state is ordered at wave vectors $\pm {\bf Q}$, and
exhibits the symmetry-breaking pattern shown in Fig.\
\ref{Orders}(b).  In general, rotation and reflection symmetries are
broken as well, due to the formation of lattice-scale currents
circulating around honeycomb plaquettes.  The specific pattern of
currents is determined by $\phi$, which within our theory is
selected spontaneously.  Consequently, one would expect a
finite-temperature Kosterlitz-Thouless transition, which has yet to be
definitively observed in a quantum Hall system.  While
graphene may eventually provide a clean setting for realizing this
physics, we note that (presumably) higher-order lattice effects will
energetically favor specific values of $\phi$.  

At filling factor $\nu = 4$, the ground state depends on the strength
of Zeeman coupling and the on-site $U$ relative to the $u_1$ and $u_2$
interactions.  The former terms favor a fully spin-polarized state,
while the latter favor a spin-singlet state with easy-plane
flavor polarization.  Variational energetics suggest that for
reasonable values of $U$ the ground state is spin polarized, and 
hence does not exhibit lattice-scale order.  

Similar physics can be expected at higher-$|n|$ Landau 
levels, although this problem has not yet been examined quantitatively.
With regard to fractional quantum Hall states, which have been studied
theoretically\cite{Peres1, KunYangSkyrmions, Toke, Apalkov, Toke2, GoerbigFractional} 
but not yet observed, the anisotropy terms
considered here should be kept in mind, particularly in situations
where a number of states are energetically competitive.

\section{Numerics}

The picture presented above for the nature of quantum Hall states at
odd-integer filling factors is supported by recent numerics by Sheng
\emph{et al}. \cite{ShengNumerics}  These authors carried out an exact
diagonalization study of the interacting Hamiltonian in Eq.\
(\ref{latticeH}), incorporating the magnetic field directly on the
lattice (\emph{i.e.}, sending $t\rightarrow t e^{i A_{\bf x x'}}$,
where $A_{\bf x x'}$ is the lattice vector potential, and including
the Zeeman energy) 
and additionally allowing for random on-site chemical
potential disorder.  Assuming the Zeeman splitting is sufficiently
large so that spin-up and spin-down Landau levels are well-separated, 
the many-body wave functions were obtained exactly upon projecting
onto the highest occupied Landau level.  Up to 24 electrons in the
projected level were considered, in systems with linear
dimensions up to $200\times 200$ and periodic boundary conditions.  

In the clean limit appropriate to the discussion above, spin- and
flavor-polarized ground states are found numerically at filling 
factors $\nu = -1$ and $\nu = 3$.  In particular, a clear easy-axis
anisotropy is observed at $\nu = -1$, corresponding to the lattice-scale 
CDW order of Fig.\ \ref{Orders}(a).  The anisotropy energy characterizing
the energy difference between the easy-axis and easy-plane
polarized states (see Fig.\ \ref{FlavorSphere}) is indeed
found to be quite small, on the order of ${\mathcal
  E}_{C}(a_0/\ell_B)^2$ as expected from our Hartree-Fock analysis.
Again, we note that this is down by a factor of $a_0/\ell_B$ compared
to ${\mathcal E}_{\rm short-range}$ since a local $\rho_{\rm stag}^2$
interaction does not split the ordering degeneracy.  
Unfortunately, the smallness of the anisotropy energy did not permit 
numerical resolution of the correspondingly small 
gap $\Delta_f$ for flavor-waves out of the easy-axis state.  
At $\nu = 3$, easy-plane flavor polarization emerges.  A strong
finite-size effect supports the lattice-scale structure predicted for
this state [see Fig.\ \ref{Orders}(b)].  Specifically, the 
easy-plane state occurs only if both linear dimensions of the system
are divisible by 3; otherwise easy-axis polarization occurs, since the
ordering at wave vectors $\pm {\bf Q}$ would then be ``frustrated''
at the boundaries.  Finite-size scaling suggests that the easy-plane
state wins in the thermodynamic limit, since the ground-state energy
per particle exhibits an upturn when the easy-axis state occurs.
Thus, numerics are entirely consistent with our analytic predictions
for the clean limit, and, importantly, can provide valuable
information about the stability of quantum Hall ferromagnetism when
disorder is included, as we will mention briefly at the end of this review.

\section{Disorder effects at $\nu = \pm 1$}

We now return to filling factors $\nu = \pm 1$, where the
flavor symmetry breaking encoded in the interacting lattice model we have
been focusing on was found to be rather weak.  Abanin
\emph{et al}.\ \cite{AbaninDisorder} have proposed an interesting 
alternative effect which
may provide the dominant flavor symmetry breaking
mechanism---\emph{disorder}.  Real graphene sheets inevitably exhibit
some degree of structural imperfection, which may arise for instance
from the presence of a substrate.  
Atomic force microscopy on graphene samples 
examined in Ref.\ \onlinecite{DisorderExpt} revealed surface 
ripples with a typical
lateral length $\xi$ of a few tens of nanometers (though ensuing
fabrication advances produced higher-mobility samples with ripples
below the resolution of their microscope \cite{DisorderExpt}).  Such 
distortions give rise to spatially varying 
electron hopping strengths, which appear as an effective
inhomogeneous magnetic field in the continuum, directed oppositely for
the two flavors.  In our notation this corresponds to sending
\begin{equation}
  D_j \rightarrow D_j -i(e/\hbar) a_j({\bf x})\tau^z  
\end{equation}
in Eq.\ (\ref{H0}), where $\vec{\delta h}({\bf x})= \nabla \times {\bm
a}$ is the effective field, whose magnitude was estimated to be 
around 0.1 to 1 T in Ref.\ \onlinecite{DisorderExpt}.  
The vector potential ${\bm a}$,
being proportional to the local hopping strength deviations, was
assumed to have white noise correlations with a correlation length
$\xi$ in Ref.\ \onlinecite{AbaninDisorder}.  

One possible consequence of this random field is that the system may
break up into domains with local easy-axis flavor polarization 
(\emph{i.e.}, local CDW order) specified by the direction of 
$\vec{\delta h}$, at the cost of domain wall energy.  This is the
Larkin-Imry-Ma state \cite{Larkin,ImryMa}.  
Abanin \emph{et al}., however, argue
that an easy-plane flavor-polarized state is energetically more 
favorable, at least for weak disorder.  The physics behind this
easy-plane selection can be understood by analogy to a ferromagnet,
where spins preferentially align \emph{perpendicular} to an applied field so
they they can effectively gain energy by relaxing toward it.  Just as 
we discussed at $\nu = 3$ and 5, the easy-plane state here 
will exhibit a Kosterlitz-Thouless transition; the transition
temperature was estimated to be a few Kelvin in Ref.\
\onlinecite{AbaninDisorder}. The
easy-plane anisotropy arising from disorder was estimated to be 
\begin{equation}
  \Delta_{\rm disorder} \sim \bigg{(}\frac{\delta
  h}{B}\bigg{)}^2\bigg{(}\frac{\xi}{\ell_B}\bigg{)}^2{\mathcal E}_C
  \sim \frac{(\delta h[{\rm T}] \xi[{\rm nm}])^2}{10\sqrt{B[{\rm T}]}}
  {\rm K},
\end{equation}
where $\delta h[{\rm T}]$ is the typical effective field strength in Teslas
and $\xi[{\rm nm}]$ is the disorder correlation length in nanometers.
With $\delta h$ in the range of 0.1 to 1 T and $\xi \sim 30$ nm, this may
indeed significantly exceed the intrinsic anisotropy $\Delta_f$ 
arising from interactions.  For systems with less structural
imperfection \cite{DisorderExpt}, however, the competition between disorder and
interactions may be quite delicate.

\section{Concluding Remarks}

We conclude by discussing the current experimental situation regarding
the physics of quantum Hall ferromagnetism in graphene.  Although
$\nu = \pm 4$ plateaus have been observed at high magnetic fields
($B \gtrsim 20$ T), \cite{HighFieldExpt,HighFieldExpt2} 
quantum Hall ferromagnetism is apparently not
yet realized in the $n = 1$ Landau level.  Plateaus at $\nu = 3, 5$
have yet to be resolved, and the activation 
energy at $\nu = 4$ was found to be dominated by Zeeman splitting 
of the spin-up
and spin-down states, rather than Coulomb energy as would be
expected for a clean system.  The situation is entirely different
for the $n = 0$ Landau level, where $\nu = 0$ and $\pm 1$
quantum Hall states have all been observed
\cite{HighFieldExpt,HighFieldExpt2}.  Furthermore, the
activation energy measured at $\nu = 1$ was several times
larger than that at $\nu = \pm 4$ and found to scale with $\sqrt{B}$,
suggesting Coulomb exchange as the origin of the gap
\cite{HighFieldExpt2}.  
Thus current
experimental graphene samples appear sufficiently clean to enable
observation of quantum Hall ferromagnetism in the $n = 0$ level, but
not at other levels.  It is worth emphasizing that the greater
stability of ferromagnetism to disorder in the $n = 0$ level is 
fully consistent with theory.  Nomura and MacDonald derived a Stoner 
criterion for the onset of interaction-driven quantum Hall states, 
which predicts a critical mobility for spontaneous
symmetry breaking that is several times larger in the $n = 1$ 
level than in the $n = 0$ level \cite{GrapheneMacDonald}.  
More recently, the numerical
study by Sheng \emph{et al}.\ \cite{ShengNumerics} 
finds a critical on-site disorder 
strength for observing a plateau at $\nu = 3$ to be roughly one 
third that at $\nu = 1$.  

Resolving the more detailed symmetry-breaking order inherent in the
likely exchange-driven $\nu = 0$ and $\pm 1$ quantum Hall 
states remains an open experimental challenge.  Ultimately, local
probes such as STM and NMR measurements would be ideal for detecting
possible microscopic patterns such as those in Fig.\
\ref{Orders}, though such experiments may be difficult since
the order would involve only a small fraction of the total number of
electrons.  Such symmetry breaking may also be revealed by
possible concomitant lattice distortions. 
On a cruder level, the qualitative trend in the
activation energy as a function of field might provide some clues as
to the nature and origin of these states.  At $\nu = \pm 1$ for
instance, the intrinsic anisotropy due to interactions scales as
$B^{3/2}$, while the anisotropy due to disorder predicted by Abanin
\emph{et al}.\ scales as $B^{-1/2}$.  Consequently, the energy for
skyrmions relative to those in the SU(4)-symmetric case should 
increase with field in the former case, but decrease in the latter,
which may be revealed through measurements of the activation energy.

\emph{Acknowledgments.} --- 
We would like to thank Leon Balents, Allan MacDonald, Kun Yang, and
Philip Kim for stimulating discussions.  
This work was supported by the National Science Foundation
through grants PHY-9907949 (M.\ P.\ A.\ F.) and 
DMR-0210790 (J.\ A.\ and M.\ P.\ A.\ F.).


\end{document}